\documentclass{iaus}

\title[short title of paper] 
{A CO, HCN and CI line survey of ULIRGs}

\author[short author list]   
{Padelis P. Papadopoulos$^1$}

\affiliation{$^1$Institute for Astronomy, ETH Zurich, Switzerland \break email: papadop@phys.ethz.ch}

\pubyear{2005}
\volume{231}  
\pagerange{119--126}
\date{?? and in revised form ??}
\jname{Proceedings Title IAU Symposium}
\editors{Dariusz C. Lis, Geoffrey A. Blake \& Eric Herbst, eds.}
\begin{document}

\maketitle

\begin{abstract}
  
  Preliminary results from a sensitive  survey of the CO J=1--0, 2--1,
  3--2, 4--3, 6--5,  HCN J=1--0, 3--2, 4--3, and CI  J=1--0 lines of a
  sample of 30 Ultraluminous Infrared Galaxies (ULIRGs) are presented.
  These reveal a tandalizing picture of the physical conditions of the
  molecular gas in  these extraordinary galaxies ($\rm L_{FIR}>10^{12}
  L_{\odot }$), with a diffuse phase dominating the low-J CO lines and
  a much denser and warmer phase dominating the CO 4-3 and 6-5 and all
  the HCN  lines. Preliminary analysis  finds the bulk of  their dense
  ($\rm n\geq 10^5\ cm^{-3}$) H$_2$ gas reservoir to be also warm $\rm
  T_k\sim (60-70) K$, without  any significant reservoir of {\it dense
    and cold}  H$_2$ gas.   This points  to most of  the dense  gas in
  ULIRGs as the immediate  ``fuel'' of their prodigious star formation
  (SF) rates, without any SF-idle dense gas reservoir present in these
  rapidly  evolving  merger/starbursts.  The  CI  J=1--0 emission  was
  found to be a robust tracer  of their total molecular gas mass under
  a large  range of physical  conditions, a potent alternative  to the
  much  weaker   emission  from  the  $   ^{13}$CO  isotopologue,  and
  especially promising as an H$_2$  tracer for similar objects at high
  redshifts.

\end{abstract}

\end{document}